\documentclass[preprintnumbers,showpacs,amsmath,amssymb,floatfix,prd,onecolumn,superscriptaddress,nofootinbib]{revtex4}

\usepackage{amsmath}
\usepackage{graphicx}
\usepackage{epsfig}
\usepackage{bm}
\usepackage{amsfonts}
\usepackage{epstopdf}
\usepackage{bm}
\usepackage[colorlinks,linkcolor=blue,backref]{hyperref}
\usepackage{cancel}

\begin{document}
	
		\title{Spin vector deviation and the gravitational wave memory effect between two free-falling gyroscopes in the plane wave spacetime}

	\author{Ke Wang}
	\affiliation{Division of Mathematical and Theoretical Physics, Shanghai Normal University, 100 Guilin Road, Shanghai 200234,  P.R.China}

	\author{Chao-Jun Feng}
\thanks{Corresponding author}
\email{fengcj@shnu.edu.cn}
\affiliation{Division of Mathematical and Theoretical Physics, Shanghai Normal University, 100 Guilin Road, Shanghai 200234,  P.R.China}

\begin{abstract}
	In the plane wave spacetime, we find that there will be a precession angle deviation between two free-falling gyroscopes when gravitational waves passed through. This kind of  angle deviation is closely related to the well-known standard velocity memory effect. Initial conditions such as the separation velocity or displacement between the two gyroscopes will affect this angle deviation. The evolutions of the angle deviation are calculated for different cases. We find that in some extreme circumstance, the angle deviation's order of magnitude produced by a rotating compact binary source could be $10^{-14}$ rads. 
	Therefore, this memory effect caused by the gravitational wave is likely to be detected in the future.
\end{abstract}


\maketitle

\section{Introduction}

  After decades of exploration, the gravitational waves from a Binary Black Hole merger was successfully detected in 2016 \cite{lasky_detecting_2016}\cite{abbott_observation_2016}, which marked the arrival
of the era of multi-messenger astronomy.

Gravitational wave memory arises from the non-oscillating components of gravitational waves. The research on memory effect can be traced back to
1974 \cite{Zeldovich:1974gvh}. Zel{'}dovich and Polnarev claimed that the distance between a pair of test masses should be changed by gravitational waves. Later studies
have shown that gravitational waves can also cause the relative velocity change between the two test masses \cite{grishchuk_gravitationalwave_1989}. These early discovered
memory effects are now usually considered as ordinary memories, since they are all produced by the final state of the gravitational wave source.
In 1991, Christodoulou et al. found that the effective energy radiated in gravitational waves would also produce memory effects, which are generally
referred to as nonlinear memory or null memory of gravitational waves \cite{christodoulou_nonlinear_1991}\cite{wiseman_christodoulous_1991}. Later calculations under PN approximation \cite{favata_post-newtonian_2009}\cite{blanchet_gravitational_2014}\cite{favata_nonlinear_2009} and numerical relativity
\cite{mitman_computation_2021}\cite{favata_gravitational-wave_2010} also support the existence of null memory effect.

The essence of null memory is that the passage of gravitational waves (or 'soft gravitons') causes a permanent change in space-time, which
will generate various observable phenomena, or the manifestation of gravitational memory. At present, a large number of manifestations of gravitational
memory have been predicted by theoretical calculation, among which the most classic and most studied is known as displacement memory \cite{favata_gravitational-wave_2010}\cite{zhang_soft_2017}, which is
described as the permanent change of distance between a pair of free-falling test particles after gravitational waves pass through. Other memory
effects include velocity memory \cite{grishchuk_gravitationalwave_1989}\cite{divakarla_first-order_2021}, spin memory \cite{pasterski_new_2016}\cite{nichols_spin_2017}, center-of-mass memory \cite{nichols_center--mass_2018}, gyroscope memory \cite{seraj_gyroscopic_2021} and so on. The purpose of this paper is to describe
a new manifestation of gravitational memory, which is closely related to the velocity memory effect.

A more profound perspective on these memory effects is the correlation with the symmetry group of asymptotically flat spacetime, the Bondi-Metzner-Sachs
group \cite{flanagan_conserved_2017}\cite{sachs_asymptotic_1962}. strominger et al. used an infrared triangle to elaborate the corresponding relationship between memory effects and the soft theorem and
the symmetries of null infinity \cite{strominger_gravitational_2016}\cite{strominger_lectures_2018}. In recent years, numerous studies on BMS groups and their extensions have linked various memory effects to various
asymptotic conserved charges \cite{nichols_spin_2017}\cite{strominger_lectures_2018}\cite{seraj_precession_2022}\cite{mao_more_2019}. These insights into gravity symmetry may open the way to the quantization of gravity.

We shall work in nonlinear plane gravitational wave spacetimes although the memory effects are frequently studied in asymptotically flat spacetimes.
As a local approximation of gravitational waves far from the source, plane waves retain most of the local properties of gravitational waves and are
formally easier to calculate. The main research objects in this paper are a pair of separate test gyroscopes, which are placed far away from the
source and their back reaction to gravitational waves is negligible. By a natural way of comparing two separated gyroscopes, we express the deviation
of the two gyroscopes as an angle, and this angle deviation is a motion independent invariant in flat spacetime. Through direct calculation, we prove
that gravitational waves will permanently change the angle deviation between the two gyroscopes, that is, the gravitational memory effects. Remarkably,
in plane gravitational wave spacetimes, the angle deviation between gyroscopes naturally corresponds to the velocity memory effect. This peculiarity
may motivate some novel proposal for detecting gravitational memory. Moreover, as a by-product of our calculation process, we will discuss the dynamics
of a single gyroscope in nonlinear plane wave spacetime and compare it to the known gyroscopic memory in asymptotic flat spacetime \cite{seraj_precession_2022}, which might
be helpful to delimit the applicable scope of plane gravitational waves.

The paper is organized as follows: Section \ref{2} reviews the generally formalism for exact plane gravitational wave spacetimes, and the geodesic motion
and memory effect in it. In Section \ref{3}, we first describe the kinematics of gyroscopes in plane wave spacetime, and then we give the precession
equations of the free-falling gyroscope with respect to a local tetrad, and describe the differences from the results in asymptotically flat spacetime.
Next, We turn to discuss two separated gyroscopes. After illustrating how two observers at different locations with different speeds can compare the gyroscopes they carry, 
we display the permanent angle deviation between the two gyroscopes generating by gravitational waves. At the end of Section \ref{3},
we discuss the effects of separation distance and separation velocity on the deviation Angle respectively under a toy model.
Finally, in Section \ref{4}, we estimate the amplitude of angle deviation generated by compact binary gravitational wave sources.

\textbf{Notation:}
Throughout this paper, we use geometric units in which $c=G=1$. Greek letters $(\mu,\nu,...)$ denote the spacetime indices and the range of available values is 
$(0,1,2,3)$, Latin indices $(a,b,...)$ denote the coordinates indices of the plane of vibration, and the available values is $(2,3)$. Hatted letters $(\hat{\mu},\hat{\nu},...)$ 
are internal Lorentz indices associated with a local frame established in section \ref{3.1}. Capital bold letters represent vectors or matrices and will 
be explained again after it appears. All the dots above the letters in this paper represent the derivative of \(U\) unless otherwise stated.
\label{Notaion}

\section{Exact plane gravitational wave Spacetimes}
\label{2}

This section briefly reviews the general plane-wave spacetime metric and its geodesic motion and memory effects.

\subsection{Metric}
\label{2.1}

Plane gravitational wave spacetime is commonly described in BJR (Baldwin-Jeffery-Rosen) coordinates and Brinkmann coordinates\cite{zhang_soft_2017}\cite{harte_optics_2015}. The general
form of metric in BJR coordinates is
\begin{equation}
	\label{bjrm}
	g=a_{i j}d x^i d x^j+2du dv,
\end{equation}
where the matrix \(\pmb{a}\) with component \(a_{i j}(u)\) are symmetric and positive. In this form, the familiar linearized transverse traceless
(TT) gauge can be expressed as \(a_{i j}=\delta _{i j}+h_{i j}^{\text{TT}}\), \cite{zhang_soft_2017} shows that the BJR coordinates have coordinate singularities,
which is typically not global, while the Brinkmann coordinates are the global coordinates of plane wave spacetime, and the metric is written as 
\begin{equation}
	\label{brm}
	g=2dU dV+\delta _{i j}d X^i d X^j+D d U^2,
\end{equation}
where D is a scalar function of \(\left(U,X^1,X^2\right)\) with the form of
\begin{equation}
	\label{DKXX}
	D=K_{i j}(U)X^iX^j=\frac{1}{2}A_{+}(U)\left(\left(X^1\right)^2-\left(X^2\right)^2\right)+A_{\times }(U)X^1X^2,
\end{equation}
where \(A_{+}(U)\) and \(A_{\times }(U)\) are the amplitude of the \(+\) and $\times $ polarization state, the non-zero components of the
Riemann curvature tensor are \(R_{i u j u}=-K_{i j}(U)\), and the only non-vanishing components of Ricci tensor is \(R_{u u}=-K_{22}-K_{11}=-\text{Tr}(K)\).
Therefore, when K is traceless, the metric in Brinkmann coordinates strictly satisfies the vacuum Einstein field equation, that is, the space-time
is Ricci flat.

There is a conversion relationship between the two coordinates \eqref{bjrm} and \eqref{brm}, which can be expressed as
\begin{equation}
	\label{trans}
	\begin{cases}
	\begin{array}{c}
	 \pmb{X}=\pmb{P(u)x} \\
	 U=u \\
	\end{array}
	 \\
	 V=v-\frac{1}{4}\pmb{x\cdot \dot{a}}\pmb{(u)\cdot x}
	\end{cases}
	\ \text{with}\qquad\left\{\pmb{
	\begin{array}{c}
	 a=P^TP \\
	 \ddot{P}=K P\\
	P^T\dot{P}=\dot{P}^TP \\
	\end{array}
	}\right.
\end{equation}
where bold \(\pmb{ X}\) and \( \pmb{ x}\) represents column vectors composed of two coordinates of the plane of vibration, and all the other bold-types represent
the 2$\times $2 matrix. Note that to get an explicit coordinate transformation you still have to solve a second-order ODE of \(\pmb{ P}\), which is \(\pmb{\ddot{P}=K
P}\). But many interesting properties can still be analyzed using the above form \cite{zhang_soft_2017}. Since \(\pmb{ P}\) has one extra degree of freedom, the coordinate transformation
is not a one-to-one mapping if the initial value of \(\pmb{ P}\) is not chosen \cite{harte_optics_2015}.

In fact, the Brinkmann coordinate \eqref{brm} can be thought as a local Lorentz frame of plane wave spacetime \cite{divakarla_first-order_2021}, which makes the geodesic equation to some
extent equivalent to the observation effect of the origin observer. Therefore, unlike the BJR coordinates, the geodesic motion under the Brinkmann
coordinates shows the abundant observation effect of gravitational waves. Moreover, Brinkmann coordinates also have computational advantages over
BJR coordinates.

The following calculation in this paper will be mainly use Brinkmann coordinates. The general geodesic motion in Brinkmann coordinates will be reviewed
below.

\subsection{General geodesic motion}
\label{2.2}

We will briefly rewrite the general solutions of geodesic equation given by \cite{flanagan_persistent_2020}, but in a more concise form. first we consider an arbitrary geodesic
$\Pi (\tau )$ with tangent vector \(\pmb{ u}\), the geodesic equation for coordinate \(U\) is given by
\begin{equation}
\label{geoU}
  \frac{d^2U}{d \tau^2}=0,
\end{equation}
with initial value \(U(\tau_0)=U_0\) the general solution can be written as 
\begin{equation}
\label{geoUs}
    U=\gamma (\tau -\tau_0)+U_0,
\end{equation}
where the parameter $\gamma $ is a conserved constant along the geodesic $\Pi $($\tau $). If we use \(\pmb{l}=\partial _V\) to represent the normal vector
of the null hypersurface parameterized by \(U\), then \(\gamma =\pmb{u}\cdot \pmb{l}\). Since there is a  linear relationship between \(U\) and $\tau$, 
we will use the coordinate \(U\) instead of the geodesic affine parameter $\tau $ in the rest of the paper. Then, the geodesic equations of the remaining 
three coordinates were written as follows:
\begin{equation}
    \label{geoX}
   \ddot{\pmb{X}}=\pmb{K}(U)\pmb{X},
\end{equation}
\begin{equation}
    \label{geoV}
    \ddot{V}=-\frac{1}{2}\dot{\pmb{D}}+2 \dot{\pmb{X}^T}\pmb{K} \pmb{X}.
\end{equation}
Hereafter the dot denotes the derivative with respect to \(U\). After setting the initial value of position as \(x^{\mu }(U_0) = (U_0,V_0, \pmb{X}_0)\),
and the initial 4-velocity as \(u^{\mu }(U_0)=\gamma \left(1, \dot{V_0},\dot{\pmb{X}_0}\right)\), the
general solution is then obtained as \footnote{Note that solution \eqref{geoXs}\eqref{geoVs} is only a formal solution, 
because \eqref{PHeq} is still an unavoidable Sturm-Liouville problem about second-order ODE \cite{zhang_sturm-liouville_2018}.}
\begin{equation}
    \label{geoXs}
    \pmb{X}=\pmb{P}(U) \pmb{X}_0+(U-U_0)\pmb{H}(U) \dot{\pmb{X}_0},
\end{equation}
\begin{equation}
    \label{geoVs}
    V=V_0-\frac{1}{2}\left(\pmb{X }^T\dot{\pmb{X}}-\pmb{X}_0 ^T\dot{\pmb{X}_0}+\frac{1}{\gamma }(U-U_0)\right),
\end{equation}
where  \(\pmb{ P}\) and \(\pmb{ H}\) are both $2\times 2 $ matrices and they satisfy the following equations respectively
\begin{equation}
    \label{PHeq}
    \pmb{\ddot{P}}=\pmb{KP},\qquad
    (U-U_0)\ddot{\pmb{H}}+2\dot{\pmb{H}}=(U-U_0)\pmb{K}\pmb{H},
\end{equation}
with the boundary conditions \(\pmb{P}(U_0)=\pmb{H}(U_0)=\pmb{I}\), \(\dot{\pmb{P}}(U_0)=\dot{\pmb{H}}(U_0)=0\). It is obviously that \(\dot{V}_0\) does not appear in the general solution \eqref{geoXs} \eqref{geoVs}, since it is just cleverly hidden in
the constant $\gamma $. This  also means that in the Brinkmann coordinate, the velocity of the gravitational wave propagation direction has no effect
on the motion of the vibration plane. So without losing generality, the following discussions will not consider the velocity of propagation direction,
but only the two-dimensional motion of the vibration plane.

By the solution \eqref{geoXs} \eqref{geoVs} we can easily write the 4-velocity for general geodesic motion as
\begin{equation}
    \label{generalu}
    \pmb{u}=\gamma \left(\partial _U-\frac{1}{2}\left[\frac{1}{\gamma }+\dot{\pmb{X }^T}\dot{\pmb{X}}+\pmb{X }^T\pmb{K}\pmb{ }\pmb{X}\right]\partial _V+\left(\dot{\pmb{X}}\right)^i\partial
_{ X^i}\right),
\end{equation}
where
\begin{equation}
    \label{generalXdot}
    \dot{\pmb{X}}=\pmb{\dot{P}X_0}+\left((U-U_0)\pmb{\dot{H}}+\pmb{H}\right)\pmb{\dot{X}_0},
\end{equation}

Thanks to the properties of local Lorentz gauge, the geodesic equation \eqref{geoU}\eqref{geoX}\eqref{geoV} in Brinkmann coordinates has the same form as the geodesic deviation equation
for the central geodesic, thus the solution \eqref{geoUs}\eqref{geoXs}\eqref{geoVs} is also the solution of the geodesic deviation equation. which makes it convenient for us to give
the expressions of displacement memory and velocity memory. Before that, let us briefly review the memory effect in plane wave spacetime.

\subsection{Memory effect}
\label{2.3}

We follow the common analytical methods, consider sandwich waves, that is, gravitational waves exist only for a zone like \(U\in \left[U_i,U_f\right]\), while
the space-time of \(U<U_i\) and \(U>U_f\) are flat but not equivalent. The theoretical foundation of non-equivalence lies in the fact that the flat
condition \(R_{\mu \nu \alpha \beta }=0\) does not constrain the linear evolutionary process, or \(K_{i j}=0\) only shows \(\pmb{P}(U)=\pmb{P}^0+U \pmb{P}^1\). In
linear theory, it can be expressed as
\begin{equation}
\label{eq2.31}
R_{\mu \nu \alpha \beta }=0\quad\Rightarrow\quad h_{i j}(U)=h_{i j}^0+U h_{i j}^1,
\end{equation}
where \(h_{i j}^0\) and \(h_{i j}^1\) are constants. Of course, this only shows that the space-time before and after the zone  can be theoretically
unequal. By using the Hamilton-Jacobi method in BJR coordinates, it shows that the space-time before and after the gravitational waves zone
are indeed not equivalent \cite{zhang_soft_2017}. In short, if we set \(a_{i j}\left(U<U_i\right)=\delta _{i j}\), we can get
\begin{equation}
\label{eq2.32}
	a_{i j}\left(U>U_f\right)\approx \delta _{i j}+2\int _{U_i}^{U_f}du'\int _{U_i}^{u'}R_{i 0 j 0}(u'')du'
\end{equation}
at the linear level. This is the physical reality of the gravitational wave memory effect. The space-time have changed permanently after the pulse
passed. This inequivalence of flat space-time will be discussed again in Section \ref{3}.

Consider a static test particle with initial coordinates \(\pmb{ X }_0\) and using the solutions \eqref{geoXs} of the geodesic equations, the change in position
after the passage of a gravitational wave can be written as
\begin{equation}
    \label{eq2.33}
    \Delta\pmb{X}=\left(\pmb{P}\left(U_f\right)-\pmb{I}\right)\pmb{X}_0,
\end{equation}
which is a general form of displacement memory in our notation. There is more information about velocity memory, since the initial position $\pmb{X}_0$ 
and the initial velocity \(\dot{ \pmb{X}_0}\) can be both taken into account simultaneously. The change of velocity can also be written as
\begin{equation}
    \label{eq2.34}
    \Delta\pmb{ V}=\dot{\pmb{P}}\pmb{X}_0+\left(\left(U_f-U_0\right)\dot{\pmb{H}}+\pmb{H}-\pmb{I}\right)\dot{\pmb{X}_0}.
\end{equation}
which is a general form of velocity memory in our notation. See the previous section \eqref{PHeq} for the definitions of matrix \(\pmb{P}\) and \(\pmb{H}\).

Of course, the manifestations of memory effect are far more than the above two. Looking for more manifestations of memory effect will provide additional
possible methods for detecting gravitational waves. In section \ref{3}, we will try to find a new memory effect by studying the separated spin
vectors.

\section{Spin vectors precession deviation}
\label{3}

This section focuses on the possible effects of the evolution of the spin vector in a general plane wave spacetime. The first step is to construct
a local tetrad so that we can study the spin vector evolution in a three-dimensional framework. In this framework, we first discuss the precession
of a free-falling gyroscope with respect to its own internal tetrad. After that, the method of comparing two separated gyroscopes is described, which
is then used to study the precession difference between the separate gyroscopes due to gravitational waves, and finally our results are discussed
under a simple gravitational wave model.

\subsection{Spin vector evolution of gyroscopes}
\label{3.1}

Consider an observer with four-velocity \(\pmb{u}=u^{\nu }\partial _{\nu }\) carries a gyroscope and falls freely, the gyroscopic spin vector $\pmb{S}$
obeys Fermi-Walker transport \((\pmb{u}\cdot \nabla \pmb{S})^{\mu }=\left(u^{\mu }a^{\nu }-u^{\nu }a^{\mu }\right)S_{\nu }\), and always satisfy
\(\pmb{S\cdot u}=0\). If we construct a local tetrad \(\left\{\pmb{e}_{\hat{\mu }}\right\}\) of the observer by making \(\pmb{e}_{\hat{0}}=\pmb{u}\),
and \(\pmb{e}_{\hat{\mu }}\cdot \pmb{e}_{\hat{\nu }}=\eta _{\hat{\mu }\hat{\nu }}\). then the spin vector is purely spatial in the tetrad, and the
precession equation of the three spatial components \(S^{\hat{i}}=\pmb{S}\cdot \pmb{e}_{\hat{i}}\) can be deduced as \cite{seraj_gyroscopic_2021}
\begin{equation}
\label{eq1}
\frac{d S^{\hat{i}}}{d \tau }=-u^a\omega _a^{\hat{i} \hat{j}}S_{\hat{j}}=\Omega _{\hat{j}}^{\hat{i}}S^{\hat{j}}  \qquad\text{with}\qquad
\omega _a^{\hat{i} \hat{j}}=\pmb{e}_{\mu }^{\hat{i}}\nabla _a\pmb{e}^{\hat{j}\mu },
\end{equation}
where \(\pmb{\omega }^{\hat{i} \hat{j}}\) is the spin connection one-form associated with the tetrad \(\left\{\pmb{e}_{\hat{\mu }}\right\}\), and
\(\pmb{ \Omega  }\) can be consider as a two-form of angular velocity\footnote{The antisymmetric tensor \(\Omega _{\hat{j}}^{\hat{i}}\) can be dualized into a vector 
\(\Omega ^{\hat{i}}=-\frac{1}{2}\epsilon ^{\hat{i}\hat{j}\hat{k}}\Omega_{\hat{j}\hat{k}}\), then the precession equation \eqref{eq1} can read as \(\dot{\pmb{S}}=\pmb{\Omega }\times \pmb{S}\)}. To calculate \(\pmb{ \Omega  }\), we need to build a local tetrad for any time-like
observer in a plane wave spacetime. The four-velocity in Brinkmann coordinates \eqref{brm} can be generally written as\footnote{Because we are working in \(U\) \(V\) coordinates, so \(\left(v^1,v^a\right)\) here does not represent the three-dimensional velocity. The four-velocity
that includes three-dimensional velocities should be set as \(\pmb{u}=\beta \left(1-v^r,-1+v^r,v^a \right)\), which is more reasonable but the calculation
will be more complicated. According to our analysis in Section 2, the velocity in the propagation direction does not affect the motion of the plane
of vibration, so without considering the velocity in the propagation direction, that is, \(v^1=-1\), then \(v^a\) can be regarded as the two-dimensional
velocity. Still writing \(v^1\) instead of -1 here is just to keep generality.}
\begin{equation}
    \label{eq2}
\pmb{u}=\gamma \left(1,v^1,v^a \right) \qquad\text{with}\qquad \pmb{u}\cdot \pmb{u}=-1\qquad \text{and}\qquad
\gamma=\left(-D-2v^1-v^av_a\right){}^{-1/2}.   
\end{equation}

First we let \(\pmb{e}_{\hat{0}}=\pmb{u}\), then subtract the part parallel to \(\pmb{e}_{\hat{0}}\) with \(\pmb{l}=\partial _V\) to get \(\pmb{e}_{\hat{1}}\).
Since the vector \(\pmb{l}=\partial _V\) is tangent to outgoing null rays, \(\pmb{e}_{\hat{1}}\) is aligned with the propagation direction of gravitational
waves. And finally we use the Gram-Schmidt orthogonalization procedure to get \(\pmb{e}_{\hat{a}}\). The whole tetrad is
\begin{equation}
\label{tetrad}
\begin{aligned}
         \pmb{e}_{\hat{0}} &=\pmb{u}, \\
         \pmb{e}_{\hat{1}} &=\frac{1}{\gamma }\pmb{l}+\pmb{u},\\
         \pmb{e}_{\hat{a}} &=\partial _a-v^a \pmb{l}. \\
\end{aligned}
\end{equation}

Then, for the general time-like motion in Brinkmann coordinates, the spin connection can be calculated by using \eqref{eq1} and \eqref{tetrad} as follows:
\begin{equation}
    \label{spin-conn}
	\begin{aligned}
	        \omega _{\nu }^{\hat{a} \hat{b}}&=0 ,\\
	        \omega _{\nu }^{\hat{1} \hat{a}}&=-\gamma \partial _{\nu }v^a+\gamma \Gamma _{\text{$\nu$b}}^1\delta ^{a b}.
	\end{aligned}
\end{equation}
By contracting with four-velocity \(\pmb{ u }\) we can also get two nonvanishing components
\begin{equation}
    \label{anglev}
    \Omega ^{\hat{1}\hat{a}}=\gamma \left(u^{\nu }\partial _{\nu }v^a-u^{\nu }\Gamma _{\nu b}^1\delta ^{a b}\right).
\end{equation}

\subsection{Free falling observer}
\label{3.2}

Now let's consider the gyroscope precession that a free-falling observer carrying a gyroscope might observe. Recall that the precession angular velocity
\eqref{anglev} with respect to the local tetrad \eqref{tetrad}, by substituting the four-velocity of general geodesic motion \eqref{generalu}, we get $ \Omega _{\hat{j}}^{\hat{i}}=0$
(For a  more detailed calculation, see Appendix \ref{appendix}). It then follows from \eqref{eq1} that
\begin{equation}
    \label{eq3.2.2}
    \frac{d S^{\hat{i}}}{d \tau }=0.   
\end{equation}

Accordingly, the spin vector component associated with the local tetrad \eqref{tetrad} remains unchanged, which means that a free-falling observer with a gyroscope
cannot detect the gravitational waves by observing the precession of the gyroscope. 

As we known that the displacement and velocity memory effects are in the order of $\mathcal{O}(1/r)$, where $r$ denotes the distance to the source. However, the precession for the spin memory is associated with high order terms, see  \cite{seraj_precession_2022}  \cite{herrera_influence_2000} and also \cite{pasterski_new_2016}, in which it shows that the leading order of the precession of gyroscopes is in the order of $\mathcal{O}(1/r^2)$. This spin effect could be also generated by the superrotation charges \cite{nichols_spin_2017} or considered as a vacuum transition under the residual Lorentz symmetry \cite{godazgar_gravitational_2022}. In our case, there is no similar charge  or symmetry, so we keep the order of $\mathcal{O}(1/r)$ in the following calculations. Actually, 
for 
the known gravitational wave memory effect of order \(o\left(r^{-1}\right)\), such as displacement memory and velocity memory, the conclusion of
plane wave space-time coincides with that of asymptotically flat spacetime \cite{zhang_soft_2017}. Actually, as a local approximation of large \(r\), the plane gravitational wave loses some information about the non-linear general gravitational wave, see \cite{flanagan_observer_2016} for an example.

In a asymptotically flat spacetime, a static observer means he/she is static relative to the source, and he/she will observe a  velocity  precession of leading order $\mathcal{O}(1/r)$, which contains the same information as the standard displacement memory. In a plane wave spacetime, we consider a static observer that  he/she is static relative to the origin of coordinates, the results
are consistent with those in asymptotically flat spacetime (See the calculation in the appendix \ref{appendix} ). However, this  velocity precession is due to non-vanishing
acceleration, and the realistic counterpart of this acceleration is not clear, so we are not going to discuss this case but only leave the results in the appendix \ref{appendix} for completeness.


We now turn to the focus on a pair of separated gyroscopes and it will show that gravitational waves create a permanent deviation among
them. At the beginning, it is necessary to specify how two observers at different locations with different speeds can compare the gyroscopes that they carry.
Since gyroscope spin vectors are spatial for self-observer, considering that parallel transport or Poincare transformation will change the spatiality,
we want a comparison method that maintains spatiality, that is, the comparison between spatial vector and spatial vector, so that the deviation angle
can be directly calculated. A natural idea is to move them together and have the same velocity, or to make them geodesics overlap, so that the spin
vectors can be compared in the same local tetrad. But how to specify the routes to coincide is not easy for a general spacetime. In a flat spacetime, the route is simple and it 
can be arbitrary. We will show by a very brief calculation that the precession generated by acceleration in a flat spacetime is path-independent.

\subsection{Path-independent precession angle}
\label{3.3}

\begin{figure}[h]
\label{figure1}
\centering
 \includegraphics{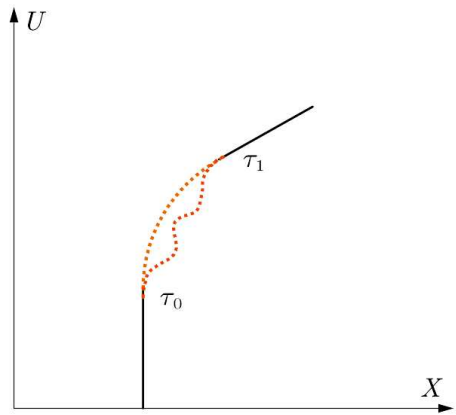}

\caption{A gyroscope moves along the worldline parameterized by $\tau$. Outside the interval $[\tau_0,\tau_1]$ is geodesic motion. In section \ref{3.3}, 
we will show that no matter what acceleration process the gyroscope undergoes, as long as the starting and ending speeds are the same, 
the precession of the gyroscope associated with a local tetrad will remain unchanged.}       
\end{figure}

Consider a worldline $\pmb{\aleph}(\tau)$ in a Minkowski space-time, make the acceleration section in $\tau \in [\tau_0,\tau_1]$, and we do not
specify the acceleration, simply write the boundary conditions as
\begin{equation}
\label{}
\frac{d\pmb{\aleph}}{d\tau }\bigg|_{\tau =\tau_0}=\pmb{u}_0, \qquad \frac{d\pmb{\aleph}}{d\tau }\bigg|_{\tau =\tau_1}=\pmb{u}.
\end{equation}
In the Minkowski spacetime,
the tetrad and the angular velocity is simply a special case of \eqref{tetrad} and \eqref{anglev}. By taking \(\pmb{K}=0\), the non-zero angular velocity reads
\begin{equation}
\label{}
\Omega _{\hat{1} \hat{a}}=-u^{\nu }\gamma \partial _{\nu }v^a=-\gamma \frac{dv^a}{d\tau }. 
\end{equation} 
The components of the angular velocity implies the plane of rotation. Without losing generality, we only consider the acceleration in the \(X^2\)
direction, then the rotation angle in the \(\pmb{e}_{\hat{1}}\wedge \pmb{e}_{\hat{2}}\) plane is obtained by integrating the angular velocity 
\(\Omega_{\hat{1} \hat{2}}\) over the interval:
\begin{equation}
    \label{prec-angle}
    \theta =-\int _{\tau 0}^{\tau 1}\gamma  \frac{dv^2}{d\tau }d\tau =-\int _{v_0^2}^{v^2}\gamma  dv^2.    
\end{equation}
It shows that the precession angle is path-independent, depending only on the starting and ending velocity. Given this property, it is clear how
two observers with different speeds on different positions can compare the gyroscopes they carry. All it takes is for them to come together. Or to
put it more precisely, make their worldlines overlap, and then we can compare them in the same tetrad. Clearly, this method is very easy to apply
in realistic scenarios. More generally, without deflection caused by force, the angle deviation of two gyroscopes is invariant as long as the spacetime remains flat no matter
how they move. While the situation in a curved spacetime is much more complicated. For the method of how to compare the spin vectors in different positions in curved
spacetime, one can refer to the affine transport method given by \cite{flanagan_observer_2016}. 

\subsection{Permanent angle deviation by separation}
\label{3.4}

\begin{figure}[h]
    \centering
    \includegraphics{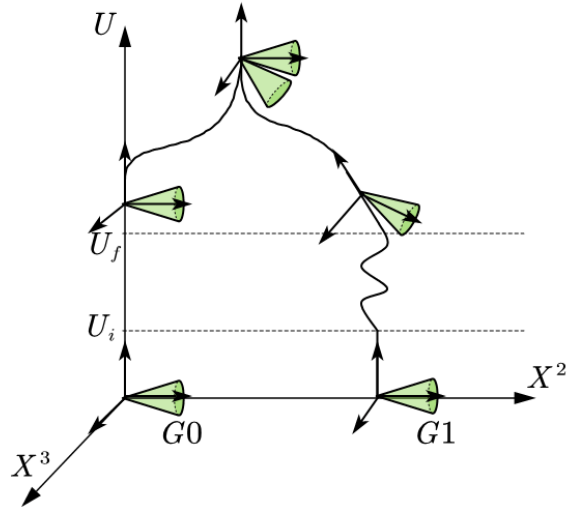}
    \caption{Two gyroscopes G1 and G0 move along their worldliness respectively in the Brinkmann coordinates. It can be regarded as the motion of G1 from the perspective of G0. 
 The region $[U_i,U_f]$ represents the spacetime location of the 
gravitational waves, while the propagation direction of waves is not shown in the diagram. The two gyrosopes are separated placed with the same orientation at the beginning. In the process of geodesic movement, the gyroscope spin vectors remain its orientation in local frame as shown in section \ref{3.2}. 
After the burst, G1 and G0 are moved together through arbitrary acceleration process, and at that time, 
the angle deviation between two gyroscopes could be observed.}
\label{figure2}
\end{figure}

In Figure \ref{figure2}, there are two gyroscopes G0 and G1 separated in a distance, in which  G0 is placed at the origin and G1 is placed at \(\pmb{X}_0=\left(X_0^2,X_0^3\right)\). Furthermore, their rotation direction is aligned at the beginning and then move along their geodesics respectively until the gravitational wave passes through.
Eq.\ \eqref{eq3.2.2} indicates that the spin vector component in the local tetrad remains unchanged. However, the local tetrad
of G1 is no longer aligned with G0 because of the velocity memory effect.  To compare the spin vectors of the two gyroscopes, their
worldlines should be overlapped in our method.  One of the natural choices is to accelerate G0 to the same velocity of G1 as follows. Firstly, accelerate G0 to  \(v_{\text{mem}}^2\) in the \(X^2\) direction and then to \(v_{\text{mem}}^3\) in the \(X^3\) direction, where \(v_{\text{mem}}^a\) denotes the velocity memory of G1 relative to G0 in the $X^a$ direction. According to Eq.\ \eqref{prec-angle}, the precession angle of G0 in the \(\pmb{e}_{\hat{1}}\wedge \pmb{e}_{\hat{a}}\) plane is calculated as
\begin{equation}
    \label{eq341}
    \theta ^a=-\int _0^{v_{\text{mem}}^a}( 2-(v)^2)^{-1/2}dv=-\arcsin\left(\frac{v_{\text{mem}}^a}{\sqrt{2}}\right).   
\end{equation}

In fact, Eq.\ \eqref{eq341}
is the relative precession angle deviation between two separated gyroscopes, since G1 and G0 are both in the same tetrad and their precession angles are zero and  \(\theta ^a\) respectively. As discussed before, without deflection caused by a force, the angle deviation
of two gyroscopes is invariant as long as the spacetime remains flat no matter how they move. So this angle will always exist after the wave passes
through. This is the memory effect of two separated spin vectors.

Since the angle deviation \eqref{eq341} contains only the velocity memory \(v_{\text{mem}}^a\), it contains the same information as that of the standard velocity memory. Thus,
the precession deviation between two separate spin vectors is just a pattern of manifestation of the velocity memory. Note that it is only true for
plane wave spacetime, the deviation of two separated spin vectors in general gravitational waves spacetime still needs further researches. The trick here
is that the standard velocity memory may be covered by the acceleration caused by various forces, but the precession deviation is a quantity that
is independent of the acceleration. It may be possible to use this to statically observe the velocity memory effect.

\subsection{Only the initial separation displacement}
\label{3.5}
In this subsection, we will discuss the precession deviation with initial separation displacement, and in next subsection the initial separation velocity will be considered. Suppose that a detector consists of two gyroscopes and can compare the angle deviation in real time. A reasonable consideration is that there
is only initial separation displacement between the two gyroscopes. Consider the simplest case, assuming that the initial separation is only in the \(X^2\) direction and the separation distance is \(L\). Since the initial
separation speed is zero, constant \(\gamma =1/\sqrt{2}\). The gravitational wave region is given by \(\left[U_i,U_f\right]\).  By using \eqref{eq2.34}, and \(\pmb{P}=\pmb{I}+\frac{1}{2}\pmb{h}+o\left(h^2\right)\) in  linear theory with the transverse traceless
gauge, we obtain the angle deviation at time $U_f$ from \eqref{eq341} as the following
\begin{equation}
    \label{eq351}
\Delta \theta =\arcsin \left(\frac{1}{\sqrt{2}}\dot{P}_{22}\left(U_f\right)L\right)=\frac{1}{2\sqrt{2}}\dot{h}_+\left(U_f\right)L+o\left(h^2\right),
\end{equation}
whose matrix form is represented below to avoid confusion, 
\begin{equation}
    \label{eq352}
    \left(
        \begin{array}{c}
         \Delta \theta ^2 \\
         \Delta \theta ^3 \\
        \end{array}
        \right)=\arcsin \left(\frac{1}{\sqrt{2}}\left(
        \begin{array}{cc}
         \dot{P}_{22} & \dot{P}_{23} \\
         \dot{P}_{32} & \dot{P}_{33} \\
        \end{array}
        \right)\left(
        \begin{array}{c}
         L \\
         0 \\
        \end{array}
        \right)\right)=\left(
        \begin{array}{c}
         \arcsin \left(\frac{1}{\sqrt{2}}\dot{P}_{22}\left(U_f\right)L\right) \\
         0 \\
        \end{array}
        \right).
\end{equation}

It is shown that the  angle deviation is approximately proportional to the initial separation distance L in this case. Actually, the angle deviation at any time $U$ in the region of $\in [U_i,U_f]$ could be given by
\begin{equation}
    \label{eq353}
    \Delta \theta (U)=\arcsin \left(\frac{1}{\sqrt{2}}\dot{P}_{22}(U)L\right)=\frac{1}{2\sqrt{2}}\dot{h}_+(U)L+o\left(h^2\right)\,,
\end{equation}
which we called the evolution equation of the angle deviation in the gravitational wave region in the simplest case.

To see the evolution of the deviation angle, we calculate it in a toy model of the gravitational collapse\cite{zhang_soft_2017}, in which
\begin{equation}
    \label{eq354}
    A_+(U)=\frac{1}{2}\frac{d^3\left(e^{-U^2}\right)}{d U^3}=\ddot{h}_+ +o\left(h^2\right)\,. 
\end{equation}

\begin{figure}[h]
    \centering
    \includegraphics{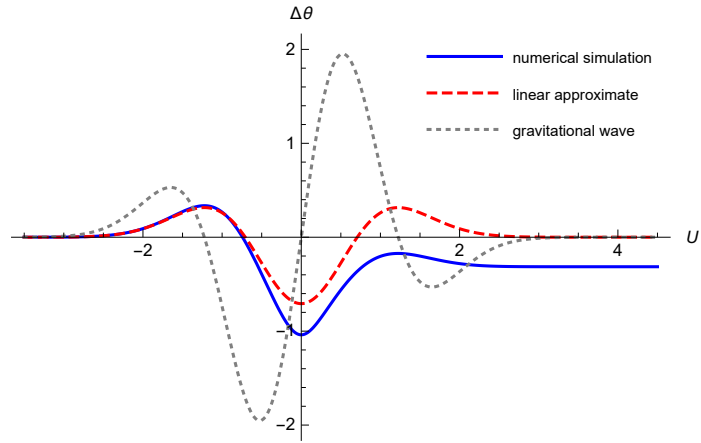}
    \caption{Evolution of deviation angle between two gyroscopes with initial separation displacement.}
      \label{figure3}
\end{figure}

The comparison between  a linear approximation and our numerical results are shown Figure \ref{figure3}. 
It shows that there is no final angle deviation for the linear approximation without any leading order velocity memory, while the numerical simulation shows that there is indeed a angle deviation when  higher order terms are taken account.

\subsection{Initial separation velocity}
\label{3.6}

Now we still consider a detector consisting of two gyroscopes, but with only initial separation velocity. The gyroscopes are all moving along their
geodesics, and we also use our method to compare the angle deviation between them in real time. We assume that  both gyroscopes G0 and G1 are at the origin at the beginning and G1 has the initial relative velocity \(v_0\)
along the \(X^2\) direction. The final velocity of G1 after the gravitational wave passes through can be readily found from \eqref{eq2.34} that
\begin{equation}
    \label{eq361}
v=\left(\left(U_f-U_0\right)\dot{H}_{22}\left(U_f\right)+H_{22}\left(U_f\right)\right)v_0
=\left(1-\frac{1}{2}h_+ +\frac{1}{2}\left(U_f-U_0\right)\dot h_++o\left(h^2\right)\right)v_0,
\end{equation}
in which we have used
\begin{equation}
	(U-U_0)\pmb H=(U-U_0)\pmb I+\frac{1}{2}(U-U_0)\pmb h-\int _{U_0}^U \pmb h dU+o\left(h^2\right).
\end{equation}

\begin{figure}[h]
    \centering   
\includegraphics{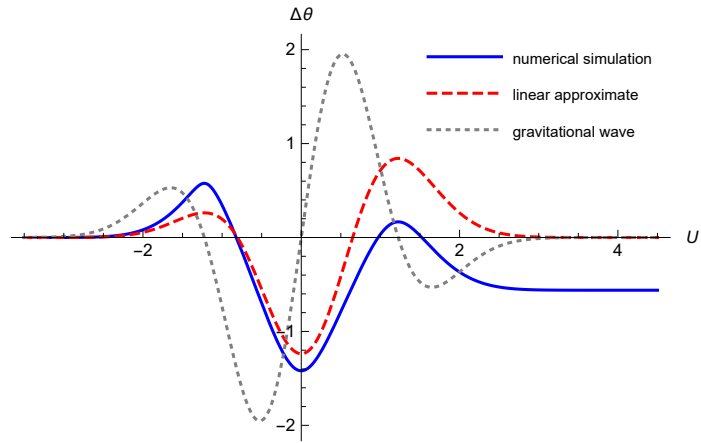}
\caption{Evolution of deviation angle between two gyroscopes with initial separation velocity.}  
 \label{figure4}  
\end{figure}

We also make G0 catch up with G1 to get the final angle deviation. By performing the same calculation as \eqref{eq353}, one can eventually finds that 
\begin{equation}
    \label{eq362}
\Delta \theta =\arcsin \left(\frac{v}{\sqrt{2}}\right)-\arcsin \left(\frac{v_0}{\sqrt{2}}\right)=-\frac{1}{2\sqrt{2}}h_+\left(U_f\right) v_0+\frac{T}{2\sqrt{2}}
\dot{h}_+\left(U_f\right)v_0+o\left(h^2\right)    ,
\end{equation}
where \(T=(U_f-U_i)\). It can be seen that the angle deviation is proportional to the initial separation velocity at the leading order. We repeat the procedure of the previous
section, first replacing \(U_f\) with \(U\) to estimate the angle deviation at anytime, and then calculate the linear approximation and numerical results
using the simple model \eqref{eq354}. The evolution of $\Delta \theta $ are shown in Figure \ref{figure4}. 

For a general case, in which there are both initial separation displacement and velocity. From the Eq.\eqref{generalXdot}, one can see  that the final velocity
has a linear relationship with the initial separation displacement and velocity. Thus, the final angle deviation at the linear level is the simple
addition of \eqref{eq351} and \eqref{eq362}, yielding
\begin{equation}
    \label{eq363}
\Delta\theta =\frac{1}{2\sqrt{2}}\left(T\dot{h}_+^{\text{mem}}-h_+^{\text{mem}}\right)v_0+\frac{1}{2\sqrt{2}}\dot{h}_+^{\text{mem}}L+o\left(h^2\right).
\end{equation}

\section{Angle deviation from compact binary sources}
\label{4}

In this section, we specifically estimate the magnitude of angle deviation memory from compact binary sources (CBS). The leading-order memory waveform
from CBS in the PN approximation is given by\cite{nichols_spin_2017}
\begin{equation}
    \label{}
h_+^{\text{mem}}=\frac{1}{48r}M \eta  x_f\sin ^2\theta \left(17+\cos ^2\theta \right)+o\left(c^{-2}\right)   \,,
\end{equation}
where M is the total mass of the binary and \(\eta =m_1 m_2/M^2\), \(x=(M\omega )^{2/3}\), where
$\omega $ is the orbital frequency and \(m_1\) \(m_2\) are the masses of two components. The initial separation of two separated gyroscopes is assumed in the same direction as the + polarization direction. Restoring our results \eqref{eq363} to
the standard unit yields
\begin{equation}
    \label{}
\Delta \theta =\frac{1}{2\sqrt{2}c}h_+^\text{mem}v_0+o\left(h^2\right)  ,  
\end{equation}
where $c$ denotes speed of light. Since the memory term \(\dot{h}_+^{\text{mem}}\) of CBS disappear, the leading order term in Eq.\eqref{eq363}
only contains the initial separation velocity \(v_0\), and the contribution from initial separation distance is hidden in the higher order terms.
Firstly, it should be noticed that there is a limitation of the plane wave approximation on the initial separation velocity. By using \(L_{\max }\) to represent the maximum separation length that allowed by the plane wave approximation and \(t_{\text{d}}\) denotes the gravitational wave duration. Then the limitation of the initial separation velocity is \(v\ll L_{\max }/t_{\text{d}}\), or else the plane wave approximation will be invalid. According to the estimation in \cite{divakarla_first-order_2021}, \(L_{\max }\) is about \(10^{17}\) meters in the case of 
\(v\ll c\). 
If we assume two gyroscopes have initial separation velocity closed to the speed
of light, then the maximum angle deviation between two gyroscopes is estimated about \(10^{-14}\) rads. In such an extreme case, it requires a very short duration of gravitational waves. For instance, two super black holes merge will emit a GW memory
burst with amplitude \(10^{-15}\) \cite{boersma_forecasts_2020}. 


%
\section{Discussion}
\label{5}

In this paper, we investigated the precession angle deviation between two separated free-falling gyroscopes in non-linear plane wave spacetime. We
first show that in a general plane-wave spacetime a free-falling gyroscopes maintaining its orientation with respect to a local tetrad \eqref{tetrad} does
not give the same results with that in the asymptotically flat spacetime \cite{seraj_gyroscopic_2021}. We then turn to discuss two separated gyroscopes. After illustrating how to
compare the local observations of two gyroscopes in a flat spacetime, we find that gravitational waves will generate a permanent angle deviation between
the two gyroscopes, and this kind of  deviation is another manifestation of the well-known standard velocity memory effect. For compact binary gravitational
wave sources, we expect to generate an angle deviation of about \(10^{-14}\) rads between two gyroscopes under the most demanding conditions.

The guiding implications of our results for gravitational wave detection warrant further discussion. What we consider here is that two gyroscopes
move along their geodesics in the gravitational wave region, and then accelerate them arbitrarily after the wave passes through to make them relatively
static. For real situations such as two separated gyroscopes placed on the ground-based detectors. For example, the test masses (mirrors) are replaced by the test gyroscopes in the detectors like  LIGO and Virgo, and the detector's control system will counteract the velocity memory and then leave a permanent angle deviation between the test gyroscopes. 
Another way to detect such effects is to let the acceleration process occur in the gravitational wave region. For example, two gyroscopes connected by a straight rod without complicated detector control system  must have experienced non-trivial acceleration processes in the gravitational wave region to remain relatively
static after the wave passed through. We will leave it to our next work.

It would be instructive to conceive of a large Sagnac interferometer \cite{sun_sagnac_1996}, which is comparable in size and technique to LIGO. Based on the strain distance now accessible to LIGO, 
a large Sagnac interferometer might be able to sense the radian change of \(10^{-25}\) rad. But given the limitations of ground-based detection, it seems more promising to resort to space-based
gravitational-wave detector, such as the future LISA \cite{amaro-seoane_laser_2017}, which could be configured as a Sagnac interferometer \cite{shaddock_operating_2004}. If such interferometers are placed at different locations in space and 
try to compare the strain differences accumulated during the passage of gravitational waves, then it is possible to observe this memory effect.

\acknowledgments
\label{6}

This work is supported by National Science Foundation of China grant No.~11105091.

%
\appendix
\label{appendix}

\section{For a free-falling observer}

Given the metric:
\begin{equation}
	g=2dU dV+\delta _{i j}d X^i d X^j+D d U^2\,,
\end{equation}
where
\begin{equation}
    D=K_{i j}(U)X^iX^j=\frac{1}{2}A_+(U)\left(\left(X^1\right)^2-\left(X^2\right)^2\right)+A_{\times }(U)X^1X^2\,,
\end{equation}
all non-zero components of affine connections and the curvature tensor are :
\begin{equation}
	\begin{split}
	    &\Gamma _{20}^1 =\Gamma _{02}^1=K_{1 a}X^a =\frac{1}{2}\partial _2D,\quad
	    \Gamma _{30}^1=\Gamma _{03}^1=K_{2 a}X^a=\frac{1}{2}\partial _3D,\\
	    &R_{0i 0 j} =-K_{i j},\quad R_{00}=-K_{22}-K_{11}.
	\end{split}
\end{equation}
The proper velocity is given by
\begin{equation}
	\pmb{u}=\gamma \left(1,v^1,v^a \right) \quad\text{with}\quad \pmb{u}\cdot \pmb{u}=-1\quad \text{and}\quad
	\gamma=\left(-D-2v^1-v^av_a\right){}^{-1/2}.
\end{equation}
Establish a local tetrad as
\begin{equation}
	\begin{aligned}
	    \pmb{e}_{\hat{0}} &=\pmb{u} ,\\
	    \pmb{e}_{\hat{1}} &=\frac{1}{\gamma }\pmb{l}+\pmb{u},\\
	    \pmb{e}_{\hat{a}} &=\partial _a-v^a \pmb{l}. \\
	\end{aligned}
\end{equation}
Since the geodesic equation of \(U\) is \(\frac{d^2U}{d\tau ^2}=0\), and $\gamma =\text{Constant}$. We firstly expand the spin connection 
\(\omega _a^{\hat{i}\hat{j}}=\pmb{e}_{\mu }^{\hat{i}}\nabla _a\pmb{e}^{\hat{j}\mu }\), yielding

\begin{equation}
\begin{split}
    	\omega _{\nu }^{\hat{i} \hat{j}}=&\pmb{e}_{\hat{i}}{}^0D\partial _{\nu }\pmb{e}^{\hat{j} 0}+\pmb{e}_{\hat{i}}{}^0\partial _{\nu }\pmb{e}^{\hat{j} 1}+\pmb{e}_{\hat{i}}{}^1\partial
    	_{\nu }\pmb{e}^{\hat{j} 0}+\pmb{e}_{\hat{i}}{}^2\partial _{\nu }\pmb{e}^{\hat{j} 2}+
    	\pmb{e}_{\hat{i}}{}^3\partial _{\nu }\pmb{e}^{\hat{j}3} \\&+\pmb{e}_{\hat{i}}{}^0\left(\Gamma _{\nu 0}^1\pmb{e}^{\hat{j} 0}+\Gamma
    	_{\nu 2}^1\pmb{e}^{\hat{j} 2}+\Gamma _{\nu 3}^1\pmb{e}^{\hat{j} 3}\right)+\pmb{e}_{\hat{i}}{}^2\Gamma _{\nu 0}^2\pmb{e}^{\hat{j} 0}+\pmb{e}_{\hat{i}}{}^3\Gamma
    	_{\nu 0}^3\pmb{e}^{\hat{j} 0}.
\end{split}	
\end{equation}
Then by using the tetrad, we obtain
\begin{equation}
	\begin{aligned}
	    \omega _{\nu }^{\hat{a} \hat{b}}&=0 ,\\
	    \omega _{\nu }^{\hat{1} \hat{a}}&=-\gamma \partial _{\nu }v^a+\gamma \Gamma _{\text{$\nu$b}}^1\delta ^{a b}.
	\end{aligned}
\end{equation}

Given the geodesic equation for \(\pmb{ X}\):
\begin{equation}
	\ddot{\pmb{X}}=\pmb{K X}=-\Gamma _{00}^a\,,
\end{equation}
the only non-zero component is
\begin{equation}
	\omega _{0 }^{\hat{1} \hat{a}}=-\gamma  \ddot{\pmb{X}}+\gamma  \Gamma _{0a}^1=\gamma \left(\Gamma _{00}^a+\Gamma _{0a}^1\right)=0\,.
\end{equation}

\section{For a relatively static observer}

In the Brinkmann coordinate, the four-velocity that relative static to the origin reads \(\pmb{u}=\frac{1}{\sqrt{2}}\left(\partial _U-\partial _V\right)\),
after the same calculation we have the tetrad:
\begin{equation}
	\begin{aligned}
	 \pmb{e}_{\hat{0}}&=\frac{1}{\sqrt{2}}(\partial _U-\partial _V) ,\\
	 \pmb{e}_{\hat{1}}&=\frac{1}{\sqrt{2}}(\partial _U-\partial _V)+\sqrt{2}\partial _V ,\\
	 \pmb{e}_{\hat{a}}&=\partial _a\,,
	\end{aligned}
\end{equation}
spin connection
\begin{equation}
	\omega _{\nu }^{\hat{1} \hat{a}}=-\pmb{e}_{\hat{1}}{}^0\partial _{\nu }v^a+\pmb{e}_{\hat{1}}{}^0\Gamma _{\nu b}^1\delta ^{ab}
	=\frac{1}{2\sqrt{2}}\partial _aD \ dU=\frac{1}{\sqrt{2}}K_{a b}X^b\ dU\,,
\end{equation}
and the angular velocity 2-form:
\begin{equation}
	\Omega _{\hat{1} \hat{a}}=-\frac{1}{2}K_{a b}X^b .
\end{equation}
The the precession angle is
\begin{equation}
	\theta =\int \pmb{\Omega }.
\end{equation}
which is actually generated by non-vanishing acceleration, and the four-acceleration is
\begin{equation}
\begin{split}
		a^{\nu }&=u^{\mu }\nabla _{\mu }u^{\nu }\pmb{=}u^{\mu }\Gamma _{\mu \alpha }^{\nu }u^{\alpha }=
		u^0\Gamma _{0\alpha }^{\nu }u^{\alpha }+u^1\Gamma _{1\alpha }^{\nu }u^{\alpha }\\
        &=\gamma ^2\Gamma _{00}^{\nu }-\gamma ^2\Gamma _{01}^{\nu }-\gamma
		^2\Gamma _{10}^{\nu }+\gamma ^2\Gamma _{11}^{\nu }=\gamma ^2\left(\Gamma _{00}^{\nu }-\Gamma _{01}^{\nu }-\Gamma _{10}^{\nu }\right),
\end{split}
\end{equation}
or 
\begin{equation}
	a^{\nu }=\frac{1}{2}\left(0,\frac{1}{2}D',-\frac{1}{2}\partial _2D,-\frac{1}{2}\partial _3D\right).
\end{equation}
%
\bibliographystyle{utphys}
\bibliography{GW-memory}

\providecommand{\href}[2]{#2}\begingroup\raggedright\begin{thebibliography}{10}

\bibitem{lasky_detecting_2016}
P.~D. Lasky, E.~Thrane, Y.~Levin, J.~Blackman, and Y.~Chen, ``Detecting
  gravitational-wave memory with {LIGO}: Implications of {GW}150914,''.
  \url{https://www.webofscience.com/wos/woscc/summary/1eb95eb2-2242-49e1-8dcb-e1f3b85519a4-49a3c6b7/date-descending/1}.
  Place: College Pk Publisher: Amer Physical Soc {WOS}:000381442800002.

\bibitem{abbott_observation_2016}
B.~Abbott, R.~Abbott, T.~Abbott, M.~Abernathy, {\em et~al.}, ``Observation of
  gravitational waves from a binary black hole merger,''.
  \url{https://link.aps.org/doi/10.1103/PhysRevLett.116.061102}.

\bibitem{Zeldovich:1974gvh}
Y.~B. Zel'dovich and A.~G. Polnarev, ``{Radiation of gravitational waves by a
  cluster of superdense stars},'' {\em Sov. Astron.} {\bfseries 18} (1974) 17.

\bibitem{grishchuk_gravitationalwave_1989}
L.~P. Grishchuk and A.~G. Polnarev, ``Gravitationalwave pulseswith
  "velocity-coded memory",''.

\bibitem{christodoulou_nonlinear_1991}
D.~Christodoulou, ``Nonlinear nature of gravitation and gravitational-wave
  experiments,''. \url{https://link.aps.org/doi/10.1103/PhysRevLett.67.1486}.

\bibitem{wiseman_christodoulous_1991}
A.~G. Wiseman and C.~M. Will, ``Christodoulou's nonlinear gravitational-wave
  memory: Evaluation in the quadrupole approximation,''.
  \url{https://link.aps.org/doi/10.1103/PhysRevD.44.R2945}.

\bibitem{favata_post-newtonian_2009}
M.~Favata, ``Post-newtonian corrections to the gravitational-wave memory for
  quasicircular, inspiralling compact binaries,''
  \href{http://arxiv.org/abs/0812.0069 [astro-ph, physics:gr-qc]}{{\ttfamily
  0812.0069 [astro-ph, physics:gr-qc]}}. \url{http://arxiv.org/abs/0812.0069}.

\bibitem{blanchet_gravitational_2014}
L.~Blanchet, ``Gravitational radiation from post-newtonian sources and
  inspiralling compact binaries,''.
  \url{http://link.springer.com/10.12942/lrr-2014-2}.

\bibitem{favata_nonlinear_2009}
M.~Favata, ``{NONLINEAR} {GRAVITATIONAL}-{WAVE} {MEMORY} {FROM} {BINARY}
  {BLACK} {HOLE} {MERGERS},''.
  \url{https://iopscience.iop.org/article/10.1088/0004-637X/696/2/L159}.

\bibitem{mitman_computation_2021}
K.~Mitman, J.~Moxon, M.~A. Scheel, S.~A. Teukolsky, M.~Boyle, N.~Deppe, L.~E.
  Kidder, and W.~Throwe, ``Computation of displacement and spin gravitational
  memory in numerical relativity.''
\newblock \url{http://arxiv.org/abs/2007.11562}.

\bibitem{favata_gravitational-wave_2010}
M.~Favata, ``The gravitational-wave memory effect,''.
  \url{https://www.webofscience.com/wos/woscc/summary/1eb95eb2-2242-49e1-8dcb-e1f3b85519a4-49a3c6b7/date-descending/1}.
  Place: Bristol Publisher: Iop Publishing Ltd {WOS}:000282154300037.

\bibitem{zhang_soft_2017}
P.-M. Zhang, C.~Duval, G.~W. Gibbons, and P.~A. Horvathy, ``Soft gravitons \&
  the memory effect for plane gravitational waves,''
  \href{http://arxiv.org/abs/1705.01378 [astro-ph, physics:gr-qc,
  physics:hep-th]}{{\ttfamily 1705.01378 [astro-ph, physics:gr-qc,
  physics:hep-th]}}. \url{http://arxiv.org/abs/1705.01378}.

\bibitem{divakarla_first-order_2021}
A.~K. Divakarla and B.~F. Whiting, ``The first-order velocity memory effect
  from compact binary coalescing sources,''
  \href{http://arxiv.org/abs/2106.05163 [gr-qc]}{{\ttfamily 2106.05163
  [gr-qc]}}. \url{http://arxiv.org/abs/2106.05163}.

\bibitem{pasterski_new_2016}
S.~Pasterski, A.~Strominger, and A.~Zhiboedov, ``New gravitational memories,''.
  \url{https://www.webofscience.com/wos/woscc/summary/1eb95eb2-2242-49e1-8dcb-e1f3b85519a4-49a3c6b7/date-descending/1}.
  Place: New York Publisher: Springer {WOS}:000399289900003.

\bibitem{nichols_spin_2017}
D.~A. Nichols, ``Spin memory effect for compact binaries in the post-newtonian
  approximation,''.
  \url{https://www.webofscience.com/wos/woscc/summary/1eb95eb2-2242-49e1-8dcb-e1f3b85519a4-49a3c6b7/date-descending/1}.
  Place: College Pk Publisher: Amer Physical Soc {WOS}:000400143100003.

\bibitem{nichols_center--mass_2018}
D.~A. Nichols, ``Center-of-mass angular momentum and memory effect in
  asymptotically flat spacetimes,''.
  \url{https://www.webofscience.com/wos/woscc/summary/1eb95eb2-2242-49e1-8dcb-e1f3b85519a4-49a3c6b7/date-descending/1}.
  Place: College Pk Publisher: Amer Physical Soc {WOS}:000444778500004.

\bibitem{seraj_gyroscopic_2021}
A.~Seraj and B.~Oblak, ``Gyroscopic gravitational memory.''
\newblock \url{http://arxiv.org/abs/2112.04535}.

\bibitem{flanagan_conserved_2017}
E.~E. Flanagan and D.~A. Nichols, ``Conserved charges of the extended
  bondi-metzner-sachs algebra,'' \href{http://arxiv.org/abs/1510.03386 [gr-qc,
  physics:hep-th]}{{\ttfamily 1510.03386 [gr-qc, physics:hep-th]}}.
  \url{http://arxiv.org/abs/1510.03386}.

\bibitem{sachs_asymptotic_1962}
R.~Sachs, ``Asymptotic symmetries in gravitational theory,''.
  \url{https://link.aps.org/doi/10.1103/PhysRev.128.2851}.

\bibitem{strominger_gravitational_2016}
A.~Strominger and A.~Zhiboedov, ``Gravitational memory, {BMS} supertranslations
  and soft theorems,''.
  \url{https://www.webofscience.com/wos/woscc/summary/1eb95eb2-2242-49e1-8dcb-e1f3b85519a4-49a3c6b7/date-descending/1}.
  Place: New York Publisher: Springer {WOS}:000368178500001.

\bibitem{strominger_lectures_2018}
A.~Strominger, ``Lectures on the infrared structure of gravity and gauge
  theory.''
\newblock \url{http://arxiv.org/abs/1703.05448}.

\bibitem{seraj_precession_2022}
A.~Seraj and B.~Oblak, ``The precession caused by gravitational waves,''
  \href{http://arxiv.org/abs/2203.16216 [gr-qc, physics:hep-th]}{{\ttfamily
  2203.16216 [gr-qc, physics:hep-th]}}. \url{http://arxiv.org/abs/2203.16216}.

\bibitem{mao_more_2019}
P.~Mao and X.~Wu, ``More on gravitational memory,''.
  \url{https://www.webofscience.com/wos/woscc/summary/1eb95eb2-2242-49e1-8dcb-e1f3b85519a4-49a3c6b7/date-descending/1}.
  Place: New York Publisher: Springer {WOS}:000467630200001.

\bibitem{harte_optics_2015}
A.~I. Harte, ``Optics in a nonlinear gravitational plane wave.''
\newblock \url{http://arxiv.org/abs/1502.03658}.

\bibitem{flanagan_persistent_2020}
E.~E. Flanagan, A.~M. Grant, A.~I. Harte, and D.~A. Nichols, ``Persistent
  gravitational wave observables: Nonlinear plane wave spacetimes,''.
  \url{https://link.aps.org/doi/10.1103/PhysRevD.101.104033}.

\bibitem{zhang_sturm-liouville_2018}
P.-M. Zhang, M.~Elbistan, G.~W. Gibbons, and P.~A. Horvathy, ``Sturm-liouville
  and carroll: at the heart of the memory effect,''
  \href{http://arxiv.org/abs/1803.09640 [gr-qc, physics:hep-th,
  physics:math-ph]}{{\ttfamily 1803.09640 [gr-qc, physics:hep-th,
  physics:math-ph]}}. \url{http://arxiv.org/abs/1803.09640}.

\bibitem{herrera_influence_2000}
L.~Herrera and J.~L.~H. Pastora, ``On the influence of gravitational radiation
  on a gyroscope,''.
  \url{https://iopscience.iop.org/article/10.1088/0264-9381/17/18/302}.

\bibitem{godazgar_gravitational_2022}
M.~Godazgar, G.~Long, and A.~Seraj, ``Gravitational memory effects and higher
  derivative actions.''
\newblock \url{http://arxiv.org/abs/2206.12339}.

\bibitem{flanagan_observer_2016}
E.~E. Flanagan and D.~A. Nichols, ``Observer dependence of angular momentum in
  general relativity and its relationship to the gravitational-wave memory
  effect.''
\newblock \url{http://arxiv.org/abs/1411.4599}.

\bibitem{boersma_forecasts_2020}
O.~M. Boersma, D.~A. Nichols, and P.~Schmidt, ``Forecasts for detecting the
  gravitational-wave memory effect with advanced {LIGO} and virgo,''.
  \url{https://www.webofscience.com/wos/woscc/summary/1eb95eb2-2242-49e1-8dcb-e1f3b85519a4-49a3c6b7/date-descending/1}.
  Place: College Pk Publisher: Amer Physical Soc {WOS}:000527519500003.

\bibitem{sun_sagnac_1996}
K.-X. Sun, M.~M. Fejer, E.~Gustafson, and R.~L. Byer, ``Sagnac interferometer
  for gravitational-wave detection,''.
  \url{https://link.aps.org/doi/10.1103/PhysRevLett.76.3053}.

\bibitem{amaro-seoane_laser_2017}
P.~Amaro-Seoane, H.~Audley, S.~Babak, {\em et~al.}, ``Laser interferometer
  space antenna.''
\newblock \url{http://arxiv.org/abs/1702.00786}.

\bibitem{shaddock_operating_2004}
D.~A. Shaddock, ``Operating {LISA} as a sagnac interferometer,''.
  \url{https://link.aps.org/doi/10.1103/PhysRevD.69.022001}.

\end{thebibliography}\endgroup

\end{document}